\documentclass[11pt]{article}
\usepackage{amssymb,amsmath,colonequals,color,multirow}
\usepackage{natbib}

\usepackage{amssymb}
\usepackage{amsfonts, amsmath, amssymb, marvosym,colonequals}
\usepackage{algorithm}
\usepackage{bm}
\usepackage{mathtools}
\usepackage{caption}
\usepackage{theorem}
\usepackage{placeins}
\usepackage{hyperref}
\usepackage{enumerate}
\allowdisplaybreaks

\newcommand{\Beta}{\mbox{\boldmath$\beta$}}

\newcommand{\bxi}{\bm{\xi}}
\newcommand{\bLambda}{\mathbf{\Lambda}}
\newcommand{\bOmega}{\mathbf{\Omega}}
\newcommand{\bPsi}{\mathbf{\Psi}}
\newcommand{\bU}{\mathbf{U}}
\newcommand{\bD}{\mathbf{D}}
\newcommand{\bT}{\mathbf{T}}
\newcommand{\bx}{\mathbf{x}}
\newcommand{\bX}{\mathbf{X}}
\newcommand{\bBeta}{\mathbf{\Beta}}

\theorembodyfont{\upshape}

\begin{document}

\title{A Latent Gaussian Mixture Model for Clustering Longitudinal Data}
\author{Vanessa S.E. Bierling and Paul D. McNicholas}
\date{\small Dept.\ of Mathematics \& Statistics, McMaster University, Hamilton, Ontario, Canada.}

\maketitle

\begin{abstract}
Finite mixture models have become a popular tool for clustering. Amongst other uses, they have been applied for clustering longitudinal data and clustering high-dimensional data. In the latter case, a latent Gaussian mixture model is sometimes used. Although there has been much work on clustering using latent variables and on clustering longitudinal data, respectively, there has been a paucity of work that combines these features. An approach is developed for clustering longitudinal data with many time points based on an extension of the mixture of common factor analyzers model. A variation of the expectation-maximization algorithm is used for parameter estimation and the Bayesian information criterion is used for model selection. The approach is illustrated using real and simulated data.\\[-10pt]

\end{abstract}


\section{Introduction}
Various work has been done on clustering high-dimensional data; including data where the number of variables $p$ is very large relative to the number of observations $n$. Some methods for clustering such data are based on the mixtures of factor analyzers model \cite[]{ghahramani97, mclachlan00a}, which assumes the variability in $p$ variables can be explained by $q<p$ latent variables for each of the $G$ components. \cite{baek10} build on work done by \cite{yoshida04,yoshida06} to introduce the closely related mixtures of common factor analyzers model. which further reduces the number of model parameters and can be useful when the mixture of factor analyzers model is not sufficiently parsimonious.

Extensive work has been done on mixture model-based clustering, where a cluster can be viewed as a component within a mixture model \cite[see][for details and discussion]{mcnicholas16a}. 
A mixture model-based approach has the advantage of ease of interpretation as well as flexibility, in that mixture models can be used for clustering various types of data and have applications in many different areas. For example, \cite{melnykov15} uses a model-based approach for clustering internet users based on the sequences of web-pages they visit, and \cite{cheam2017} use a model-based approach for clustering spatiotemporal air quality data. Various work has been done on clustering longitudinal data. For example, \cite{mcnicholas10b} introduce a family of mixture models that utilizes a covariance matrix decomposition that takes into account the relationship between measurements taken at different time points, and \cite{han2017} introduce a Bayesian semi-parametric method for clustering data that identifies the temporal pattern common to all subjects as well as temporal patterns unique to individual subjects. However, there is a dearth of work specifically for clustering high-dimensional longitudinal data, i.e., data that contain measurements taken at many time points. This paper introduces a method for clustering longitudinal data using a latent Gaussian mixture model that builds on the approaches introduced by \cite{baek10} and \cite{mcnicholas10b}. In particular, it allows the variability in measurements taken at a large number of time points, $p$, to be explained using a smaller number of time points, $q$.

\section{Methodology} \label{sec:themodel}
\subsection{Background}
The density of a $p$-dimensional random variable $\mathbf{X}$ that arises from a finite mixture model with $G$ components is of the form
\begin{equation}
f(\bx \mid \bm\vartheta) = \sum_{g=1}^G \pi_g f_g(\bx \mid \bm\theta_g),
\label{eq:mixturemodel}
\end{equation}
where $\pi_g>0$ is the $g$th mixing proportion with $\sum_{g=1}^G\pi_g=1$, $f_g(\bx \mid \bm\theta_g)$ is the $g$th component density, and $\bm\vartheta = (\pi_1, \dots, \pi_G, \bm\theta_1, \dots, \bm\theta_G)$ is the vector of parameters. 
The most commonly used mixture model is the Gaussian mixture model and 
the density of this model is of the form
\begin{equation}
f(\bx\mid \bm\vartheta) = \sum_{g=1}^G \pi_g \phi(\bx \mid \bm\mu_g, \bm\Sigma_g),
\label{eq:mixturemodel}
\end{equation}
where $\phi(\bx \mid \bm\mu_g, \bm\Sigma_g)$ is the density of the multivariate Gaussian distribution with mean vector  $p \times 1$ mean vector $\bm\mu_g$ and $p \times p$ covariance matrix $\bm\Sigma_g$, for $g=1,\dots,G$.


\cite{pourahmadi99,pourahmadi00} uses a modified Cholesky decomposition on the covariance matrix $\mathbf\Sigma$ of a $p$-dimensional random variable, which can be used to model longitudinal data. This decomposition is given by
\begin{equation}
\label{eq:cholesk}
\bT \mathbf\Sigma \bT^{'}=\bD,
\end{equation}
where $\bT$ is a unique $p \times p$ unit lower triangular matrix, the lower triangular elements of which are interpreted as generalized autoregressive parameters, and $\bD$ is a unique $p \times p$ diagonal matrix, the (strictly positive) diagonal elements of which are interpreted as innovation variances \cite[]{pourahmadi99}.
This allows the value of a measurement taken at time $t$, denoted $Y_t$, to be predicted using the measurements taken at previous time points, denoted $Y_{t-1}, \dots, Y_1$. Specifically, the linear least-squares predictor of $Y_t$ is given by
\begin{equation}
\hat{Y}_t = \mu_t + \sum_{s=1}^{t-1} (-\phi_{ts}) (Y_s-\mu_s) + \sqrt{d_t} \epsilon_t,
\label{eq:longitudinal}
\end{equation}
where the $\phi_{ts}$ are the lower triangular elements of $\bT$, the $d_t$ are the diagonal elements of $\bD$, and $\epsilon_t \sim \mathcal{N}(0, 1)$ for $t=2, \dots, p$.

To cluster longitudinal data, \cite{mcnicholas10b} use a Gaussian mixture model with each component covariance matrix decomposed according to (\ref{eq:cholesk}), i.e.,
\begin{equation}
\bT_g \mathbf\Sigma_g \bT_g^{'}=\bD_g
\end{equation}
for $g=1,\dots,G$. Equivalently, each component precision matrix is given by $$\mathbf\Sigma_g^{-1}=\bT_g^{'} \bD_g^{-1} \bT_g.$$ Imposing constraints on the $\bT_g$ and/or $\bD_g$ matrices results in a family of eight mixture models, ranging from $p(p-1)/2+1$ covariance parameters in the most constrained to $G[p(p-1)]/2 + Gp$ in the least constrained.


Suppose $p$-dimensional $\bX_1,\ldots,\bX_n$ are observed. The mixture of factor analyzers (MFA) model \cite[]{ghahramani97, mclachlan00b} assumes that $\bX_i$ can be modelled as 
\begin{equation}
\bX_i = \bm\mu_g + \bLambda_g \bU_{ig} + \bm\varepsilon_{ig}
\label{eq:MFA1}
\end{equation}
with probability $\pi_g$, for $i=1,\dots,n$ and $g=1,\dots,G$. Note that $\bm\mu_g$ is a $p\times 1$ mean vector$,\bLambda_g$ is a $p \times q$ matrix of factor loadings, $q<p$, the $q \times 1$ latent factors $\bU_{ig}$ are independent $\sim \mathcal{N} (\mathbf{0},\textbf{I}_q)$, the $\bm\varepsilon_{ig}$ are independent $\sim \mathcal{N} (\mathbf{0},\bPsi_g)$, where $\bPsi_g$ is a $p \times p$ diagonal matrix, and the $\bU_{ig}$ and $\bm\varepsilon_{ig}$ are independent of each other. It follows that the density of $\bX_i$ is given by
\begin{equation}
f(\bx_i \mid \bm\vartheta) = \sum_{g=1}^G \pi_g \phi(\bx_i \mid \bm\mu_g, \bLambda_g \bLambda_g^{'} + \bPsi_g).
\label{eq:MFA}
\end{equation}
The number of parameters in the MFA model can be further reduced by imposing certain constraints on the mean vector and covariance matrix; this leads to the mixtures of common factor analyzers (MCFA) model introduced by \cite{baek10}, which assumes $\bX_i$ can be modelled as 
\begin{equation}
\bX_i = \bLambda \bU_{ig} + \bm\varepsilon_{ig}
\label{eq:MCFA1}
\end{equation}
with probability $\pi_g$, for $i=1,\dots,n$ and $g=1,\dots,G$. Here, $\bLambda$ is a $p \times q$ matrix of factor loadings, the latent factors $\bU_{ig}$ are independent $\sim \mathcal{N} (\bxi_g,\bOmega_g)$, the $\bm\varepsilon_{ig}$ are independent $\sim \mathcal{N} (\bf{0},\bPsi)$, and the $\bU_{ig}$ and $\bm\varepsilon_{ig}$ are independent of each other. Note that $\bxi_g$ is a $q$-dimensional vector, $\bOmega_g$ is a $q \times q$ symmetric matrix, and $\bPsi$ is a $p \times p$ diagonal matrix. It follows that the density of $\bX_i$ is given by
\begin{equation}
f(\bx_i \mid \bm\vartheta) = \sum_{g=1}^G \pi_g \phi (\bx_i \mid \bLambda \bxi_g, \bLambda \bOmega_g \bLambda^{'} + \bPsi).
\label{eq:MCFA2}
\end{equation}

\subsection{The Model}
The proposed model utilizes the modified Cholesky decomposition on the latent covariance matrix in the MCFA model; that is, $\bOmega_g^{-1}=\bT_g^{'} \bD_g^{-1} \bT_g$, where $\bT_g$ is a $q \times q$ unit lower triangular matrix and $\bD_g$ is a $q \times q$ diagonal matrix. It follows that the density given in (\ref{eq:MCFA2}) can be written
\begin{equation}
f(\bx_i \mid \bm\vartheta) = \sum_{g=1}^G \pi_g \phi (\bx_i \mid \bLambda \bxi_g, \bLambda (\bT_g^{'} \bD_g^{-1} \bT_g)^{-1} \bLambda^{'} + \bPsi).
\end{equation}
Using results from \cite{baek10}, the total number of parameters in the model is 
\begin{equation}(G-1) + Gq + (pq-q^2) + p + G\left[\frac{q (q-1)}{2}\right]+Gq.
\label{eq:parameters}
\end{equation}
Similarly to the family of mixture models for longitudinal data introduced by \cite{mcnicholas10b}, the value in (\ref{eq:parameters}) can be reduced further by imposing constraints on the $\bT_g$ and/or $\bD_g$ matrices. Doing so leads to a family of eight mixture models; details of which, including the number of covariance numbers required to be estimated for each, are given in Table \ref{table:constraints}. Note that, unlike the aforementioned family of mixture models for longitudinal data, these constraints directly impact the latent space. 
\begin{table}[h!]
\centering
\caption{The nomenclature and covariance structure for each member in the family of models, together with the number of free covariance parameters in $\bOmega_g$.}
\label{table:constraints}
\begin{tabular}{llllr}
\hline
\textbf{Model} & $\bT_g$ & $\bD_g$ & $\bD_g$ & \textbf{No.\ Free Cov.\ Parameters in }$\bOmega_g$ \\ \hline
EEA & Equal & Equal & Anisotropic & $q(q-1)/2+q$ \\
VVA & Variable & Variable & Anisotropic & $G[q(q-1)/2]+Gq$ \\
VEA & Variable & Equal & Anisotropic & $G[q(q-1)/2]+q$ \\
EVA & Equal & Variable & Anisotropic & $q(q-1)/2]+Gq$ \\
VVI & Variable & Variable & Isotropic & $G[q(q-1)/2]+G$ \\
VEI & Variable & Equal & Isotropic & $G[q(q-1)/2]+1$ \\
EVI & Equal & Variable & Isotropic & $q(q-1/2)+G$ \\
EEI & Equal & Equal & Isotropic & $q(q-1/2)+1$ \\ \hline                              
\end{tabular}
\end{table}

\subsection{Model Fitting}
\label{sec:modelfitting}
The models are fitted using an EM algorithm \cite[]{dempster77, baek10}, where the $\bx_1,\ldots,\bx_n$ are taken to be the observed data and the component membership labels $z_{ig}$ and latent factors $\mathbf{u}_{ig}$ are taken to be the missing data, where $z_{ig} = 1$ if the $i$th observation belongs to the $g$th component and $z_{ig}=0$ otherwise.
The complete-data log-likelihood is given by
\begin{equation}
l_c (\bm\vartheta) = \sum_{i=1}^n \sum_{g=1}^G z_{ig} \text{log} \left[ \pi_g \phi (\mathbf{x}_i \mid \bLambda \mathbf{u}_{ig}, \bPsi) \phi (\mathbf{u}_{ig} \mid \bxi_g, (\bT_g^{'} \bD_g^{-1} \bT_g)^{-1}) \right]
\label{eq:loglikelihood}
\end{equation}
and its expected value is given by
\begin{align}
\label{eq:Q}
\begin{split}
Q = \sum_{i=1}^{n} \sum_{g=1}^{G} & \hat{z}_{ig} \bigg[ \text{log}\pi_g - p\text{log}(2\pi) + \frac{1}{2}\text{log}\left| \bT_g^{'} \bD_g^{-1} \bT_g \right| + \frac{1}{2} \text{log} \left| \bPsi^{-1} \right| 
\\ &- \frac{1}{2} \text{tr}\left\{ \bT_g^{'} \bD_g^{-1} \bT_g \mathbb{E} \lbrack (\bU_{ig}-\bxi_g)(\bU_{ig}-\bxi_g)^{'} \mid \mathbf{x}_i \rbrack \right\}
\\ &- \frac{1}{2} \text{tr} \left\{ \bPsi^{-1} \mathbb{E} \lbrack (\mathbf{X}_i - \bLambda \bU_{ig}) (\mathbf{X}_i - \bLambda \bU_{ig})^{'} \mid \mathbf{x}_i \rbrack \right\} \bigg],
\end{split}
\end{align}
where the $\hat{z}_{ig}$ are the expected values of the component membership labels, given by
\begin{equation}
\hat{z}_{ig} = \frac{\hat{\pi}_g \phi(\textbf{x}_i \mid \hat\bLambda \hat\bxi_g, \hat\bLambda (\hat\bT_g^{'} \hat\bD_g^{-1} \hat\bT_g)^{-1} \hat{\bLambda}^{'}+\hat\bPsi)}{\sum_{h=1}^{G} \hat{\pi}_h \phi(\textbf{x}_i \mid \hat\bLambda \hat\bxi_h, \hat\bLambda (\hat\bT_h^{'} \hat\bD_h^{-1} \hat\bT_h)^{-1} \hat{\bLambda}^{'}+\hat\bPsi)}.
\label{eq:z}
\end{equation}

In the E-step of the algorithm, the $\hat{z}_{ig}$ are updated according to (\ref{eq:z}).
In the M-step, the model parameter estimates, obtained by maximizing $Q$, are updated. Maximizing $Q$ with respect to $\pi_g$ yields the update
\begin{equation}
\hat{\pi}_g = \frac{n_g}{n},
\end{equation}
where $n_g=\sum_{i=1}^n \hat{z}_{ig}$. Differentiating $Q$ with respect to $\bxi_g$, $\bT_g$, $\bD_g^{-1}$, $\bLambda$, and $\bPsi^{-1}$, respectively, \citep[and using results from][]{lutkepohl96} gives the following score functions, derivations of which are included in the Appendix. Only the VVA model and its derivations are included here, but derivations of the other models in Table \ref{table:constraints} follow similarly. 
\begin{align*}
& S_{1}(\bm{\vartheta}_{s}) = \frac{\partial Q}{\partial \bxi_g} 
= \bT_{g}^{'} \bD_g^{-1} \bT_{g} \sum_{i=1}^{n} \hat{z}_{ig} \left( \mathbb{E} \lbrack \mathbf{U}_{ig} \mid \mathbf{x}_{i}, z_{ig} = 1 \rbrack - \bxi_g \right),
\\[1.5em] & S_{2}(\bm{\vartheta}_{s}) = \frac{\partial Q}{\partial \bT_{g}}
= - n_{g} \bD_g^{-1} \bT_{g} \mathbf{S}_{g},
\\[1.5em] & S_{3}(\bm{\vartheta}_{s}) = \frac{\partial Q}{\partial \bD_g^{-1}}
= \frac{n_{g}}{2} \bD_g - \frac{1}{2} \sum_{i=1}^{n} \hat{z}_{ig} \bT_{g} \Big( \mathbb{V}\text{ar} \lbrack \bU_{ig} - \bxi_g \mid \mathbf{x}_{i}, z_{ig} = 1 \rbrack \\& \hspace{4em}
\qquad\qquad\qquad\qquad+ \mathbb{E} \lbrack \bU_{ig} - \bxi_g \mid \mathbf{x}_{i}, z_{ig} = 1 \rbrack \mathbb{E} \lbrack (\bU_{ig} - \bxi_g)^{'} \mid \mathbf{x}_{i}, z_{ig} = 1 \rbrack \Big) \bT_{g}^{'},
\\[1.5em] & S_{4}(\bm{\vartheta}_{s}) = \frac{\partial Q}{\partial \bLambda} 
= \sum_{i=1}^{n} \sum_{g=1}^{G} \hat{z}_{ig} \bPsi^{-1} \left( \mathbf{x}_{i} \mathbb{E} \lbrack \bU_{ig}^{'} \mid \mathbf{x}_{i}, z_{ig} = 1 \rbrack - \bLambda \mathbb{E} \lbrack \bU_{ig} \bU_{ig}^{'} \mid \mathbf{x}_{i}, z_{ig} = 1 \rbrack \right),
\\[1.5em] & S_{5}(\bm{\vartheta}_{s}) = \frac{\partial Q}{\partial \bPsi^{-1}} 
= \frac{n}{2} \bPsi - \frac{1}{2} \sum_{i=1}^{n} \sum_{g=1}^{G} \hat{z}_{ig} \Big( \mathbf{x}_{i} \mathbf{x}_{i}^{'} - \bLambda \mathbb{E} \lbrack \bU_{ig} \mid \mathbf{x}_{i}, z_{ig} = 1 \rbrack \mathbf{x}_{i}^{'} - \mathbf{x}_{i} \mathbb{E} \lbrack \bU_{ig}^{'} \mid \mathbf{x}_{i}, z_{ig} = 1 \rbrack \bLambda^{'} \\& \hspace{4em}
\qquad\qquad\qquad\qquad+ \bLambda \mathbb{E} \lbrack \bU_{ig} \bU_{ig}^{'} \mid \mathbf{x}_{i}, z_{ig} = 1 \rbrack \bLambda^{'} \Big),
\end{align*}
{where } 
$$\mathbf{S}_{g} = \frac{1}{n_{g}} \sum_{i=1}^{n} \hat{z}_{ig} \left[ (\mathbf{I}_{q}-\hat{\bBeta}_{g} \hat{\bLambda}) (\hat{\bT}_{g}^{'} \hat{\bD}_g^{-1} \hat{\bT}_{g})^{-1} + \hat{\bBeta}_{g} ( \mathbf{x}_{i} - \hat{\bLambda} \hat{\bxi}_{g}) ( \mathbf{x}_{i} - \hat{\bLambda} \hat{\bxi}_{g})^{'} \hat{\bBeta}_{g}^{'} \right]$$ \\
{and } $$\hat{\bBeta}_{g} = (\bT_g^{'} \bD_g^{-1} \bT_g)^{-1} \bLambda^{'} (\bLambda (\bT_g^{'} \bD_g^{-1} \bT_g)^{-1} \bLambda^{'} + \bPsi)^{-1}.$$

Note that the joint distribution of $\textbf{X}_i$ and $\textbf{U}_{ig}$ given membership in the $g$th component is given by
\begin{align*}
\begin{bmatrix}
\textbf{X}_i \\
\textbf{U}_{ig}
\end{bmatrix}
\Bigg\rvert z_{ig} = 1
\sim
\begin{pmatrix}
\begin{bmatrix}
\bLambda \bxi_g \\
\bxi_g
\end{bmatrix},
\begin{bmatrix}
\bLambda (\bT_g^{'} \bD_g^{-1} \bT_g)^{-1} \bLambda^{'} + \bPsi & \bLambda (\bT_g^{'} \bD_g^{-1} \bT_g)^{-1} \\
(\bT_g^{'} \bD_g^{-1} \bT_g)^{-1} \bLambda^{'} & (\bT_g^{'} \bD_g^{-1} \bT_g)^{-1}
\end{bmatrix}
\end{pmatrix},
\end{align*}
from which it follows that
\begin{equation}
\label{eq:expectation1}
\mathbb{E} \left[ \bU_{ig} | \bx_{i}, z_{ig} = 1 \right] = \bxi_g + \Beta_g (\bx_{i} - \bLambda \bxi_g)
\end{equation}
and
\begin{align}
\label{eq:expectation2}
\begin{split}
\mathbb{E} [ \bU_{ig} &\bU_{ig}^{'} | \bx_{i}, z_{ig} = 1 ] \\
&= \mathbb{V} [\bU_{ig} | \bx_i, z_{ig}=1] + \mathbb{E} [\bU_{ig} | \bx_i, z_{ig}=1] \mathbb{E} [\bU_{ig} | \bx_i, z_{ig}=1]^{'} \\
&= (\textbf{I}_q - \Beta_g \bLambda) (\bT_g^{'} \bD_g^{-1} \bT_g)^{-1} + \left[ \bxi_g + \Beta_g (\bx_i - \bLambda \bxi_g ) \right]\left[ \bxi_g + \Beta_g (\bx_i - \bLambda \bxi_g ) \right]^{'}.
\end{split}
\end{align}
Using (\ref{eq:expectation1}) and (\ref{eq:expectation2}), the parameter estimates are obtained after setting the above score functions to the appropriate zero vector or matrix; further details are given in the Appendix. First, solving $S_1 (\hat{\bxi}_{g}^{\text{new}}, \hat{\bT}_{g}^{\text{new}}, \hat{\bD}_{g}^{\text{new}}, \hat{\bLambda}^{\text{new}}, \hat{\bPsi}^{\text{new}}) = \bf{0}$ gives
\begin{align*}
\hat{\bxi}_{g}^{\text{new}} = \hat{\bxi}_g + \frac{1}{n_g} \sum_{i=1}^{n} \hat{z}_{ig} \hat{\Beta}_{g} (\mathbf{x}_{i} - \hat{\bLambda} \hat{\bxi}_{g} ).
\end{align*}

To find $\hat{\bT}_{g}^{\text{new}}$, first denote the lower triangular elements of $\bT_g$ by $\phi_{ij}^{(g)}$ for $i=2,\dots,q$ and $j=~1,\dots,i-1$, so that
\[ \bT_g = \left( \begin{array}{cccccc}
1 & 0 & 0 & 0 & \cdots & 0 \\
\phi_{21}^{(g)} & 1 & 0 & 0 & \cdots & 0 \\
\phi_{31}^{(g)} & \phi_{32}^{(g)} & 1 & 0 & \cdots & 0 \\
\vdots & \vdots &  & \ddots &  & \vdots \\
\phi_{q-1,1}^{(g)} & \phi_{q-1,2}^{(g)} & \cdots & \phi_{q-1,q-2}^{(g)} & 1 & 0 \\
\phi_{q1}^{(g)} & \phi_{q2}^{(g)} & \cdots & \phi_{q,q-2}^{(g)} & \phi_{q,q-1}^{(g)} & 1 \end{array} \right).\]
Then, setting the lower triangular element of $S_2(\bm\vartheta_s)$ equal to 0 and solving for the $\phi_{ij}^{(g)}$ results in $q-1$ systems of linear equations, the solutions of which are given by
\[ \left( \begin{array}{c}
\hat{\phi}_{r1}^{(g)} \\
\hat{\phi}_{r2}^{(g)}\\
\vdots\\
\hat{\phi}_{r,r-1}^{(g)}\end{array} \right)
=  - \left( \begin{array}{cccc}
s_{11}^{(g)} & s_{21}^{(g)} & \cdots & s_{r-1,1}^{(g)}\\
s_{12}^{(g)} & s_{22}^{(g)} & \cdots & s_{r-1,2}^{(g)}\\
\vdots & \vdots & \ddots & \vdots\\
s_{1,r-1}^{(g)} & s_{2,r-1}^{(g)} & \cdots & s_{r-1,r-1}^{(g)}\end{array} \right)^{-1}
\left( \begin{array}{c}
s_{r1}^{(g)} \\
s_{r2}^{(g)}\\
\vdots\\
s_{r,r-1}^{(g)}\end{array} \right) \] 
for $r=2,\dots,p$, where $s_{ij}^{(g)}$ denotes the $(i,j)^{\text{th}}$ element of $\mathbf{S}_g$. Thus, the lower triangular elements of $\hat{\bT}_{g}^{\text{new}}$ are given by the $\hat{\phi}_{ij}^{(g)}$. \\

Solving $\text{diag}\left\{ S_3 (\hat{\bxi}_{g}^{\text{new}}, \hat{\bT}_{g}^{\text{new}}, \hat{\bD}_g^{\text{new}}, \hat{\bLambda}^{\text{new}}, \hat{\bPsi}^{\text{new}}) \right\} = \bf{0}$ gives
\begin{align*}
\hat{\bD}_g^{\text{new}} = \text{diag} \left\{ \hat{\bT}_{g} (\mathbf{I}_{q}-\hat{\bBeta}_{g} \hat{\bLambda}) 
\hat{\bT}_{g}^{-1} \hat{\bD}_g + \frac{1}{n_{g}} \sum_{i=1}^{n} \hat{z}_{ig} \hat{\bT}_{g} \hat{\bBeta}_{g} ( \mathbf{x}_{i} - \hat{\bLambda} \hat{\bxi}_{g}) ( \mathbf{x}_{i} - \hat{\bLambda} \hat{\bxi}_{g})^{'} \hat{\bBeta}_{g}^{'} \hat{\bT}_{g}^{'} \right\},
\end{align*}
solving $S_4 (\hat{\bxi}_{g}^{\text{new}}, \hat{\bT}_{g}^{\text{new}}, \hat{\bD}_g^{\text{new}}, \hat{\bLambda}^{\text{new}}, \hat{\bPsi}^{\text{new}}) = \bf{0}$ gives
\begin{align*}
& \hat{\bLambda}^{\text{new}} = \left\{
\sum_{i=1}^{n} \sum_{g=1}^{G} \hat{z}_{ig} \mathbf{x}_{i} \left[
\hat{\bxi}_g + \hat{\Beta}_{g} (\mathbf{x}_{i} - \hat{\bLambda} \hat{\bxi}_{g} )
\right]^{'} \right \} \Bigg\{ \sum_{i=1}^{n} \sum_{g=1}^{G} \hat{z}_{ig} (\mathbf{I}_{q} - \hat{\Beta}_{g} \hat{\bLambda}) (\hat{\bT}_{g}^{'} \hat{\bD}_g^{-1} \hat{\bT}_{g})^{-1} \\& \hspace{3.5em}
+ \sum_{i=1}^{n} \sum_{g=1}^{G} \hat{z}_{ig} \lbrack \hat{\bxi}_{g} + \hat{\bBeta}_{g} ( \mathbf{x}_{i} - \hat{\bLambda} \hat{\bxi}_{g}) \rbrack \lbrack \hat{\bxi}_{g} + \hat{\bBeta}_{g} ( \mathbf{x}_{i} - \hat{\bLambda} \hat{\bxi}_{g}) \rbrack ^{'} \Bigg\}^{-1},
\end{align*}
and solving $\text{diag}\left\{ S_5 (\hat{\bxi}_{g}^{\text{new}}, \hat{\bT}_{g}^{\text{new}}, \hat{\bD}_g^{\text{new}}, \hat{\bLambda}^{\text{new}}, \hat{\bPsi}^{\text{new}}) \right\} = \bf{0}$ gives
\begin{align*}
& \hat{\bPsi}^{\text{new}} = \frac{1}{n} \text{diag} \Bigg\{ \sum_{i=1}^{n} \sum_{g=1}^{G} \hat{z}_{ig}
\bigg( \mathbf{x}_{i} \mathbf{x}_{i}^{'} - \hat{\bLambda} \left[ \hat{\bxi}_g + \hat{\Beta}_{g} (\mathbf{x}_{i} - \hat{\bLambda} \hat{\bxi}_{g} ) \right] \mathbf{x}_{i}^{'} - \mathbf{x}_{i} \left[ \hat{\bxi}_g + \hat{\Beta}_{g} (\mathbf{x}_{i} - \hat{\bLambda} \hat{\bxi}_{g} ) \right] ^{'} \hat{\bLambda}^{'} \\& \hspace{3.5em}
+ \hat{\bLambda} \left[ (\mathbf{I}_{q} - \hat{\Beta}_{g} \hat{\bLambda})(\hat{\bT}_{g}^{'} \hat{\bD}_g^{-1} \hat{\bT}_{g} )^{-1} + \left[ \hat{\bxi}_g + \hat{\Beta}_{g} (\mathbf{x}_{i} - \hat{\bLambda} \hat{\bxi}_{g} ) \right]\left[ \hat{\bxi}_g + \hat{\Beta}_{g} (\mathbf{x}_{i} - \hat{\bLambda} \hat{\bxi}_{g} ) \right] ^{'} \right] \hat{\bLambda}^{'} \bigg) \Bigg\}.
\end{align*}

\noindent The pseudocode for the EM algorithm is then given by: 

\noindent initialize $\hat{z}_{ig}$\\
initialize $\hat{\pi}_g, \hat{\bxi}_g, \hat{\bT}_g, \hat{\bD}_g, \hat{\bLambda}_g, \hat{\bPsi}$\\
while not converged

compute $\hat{\Beta}_g$
	
update $\hat{\pi}_g, \hat{\bxi}^{\text{new}}_g, \hat{\bT}^{\text{new}}_g, \hat{\bD}^{\text{new}}_g, \hat{\bLambda}^{\text{new}}_g, \hat{\bPsi}^{\text{new}}$
	
update $\hat{z}_{ig}$

check convergence criterion

$\hat{\bxi}_g \leftarrow \hat{\bxi}^{\text{new}}_g, \hat{\bT}_g \leftarrow \hat{\bT}^{\text{new}}_g, \hat{\bD}_g \leftarrow \hat{\bD}^{\text{new}}_g, \hat{\bLambda}_g \leftarrow \hat{\bLambda}^{\text{new}}_g, \hat{\bPsi} \leftarrow \hat{\bPsi}^{\text{new}}$\\
end while

Various possible convergence criteria can be used, such as those based on Aitken's acceleration \cite[]{aitken26,bohning94,mcnicholas10a}. For the simulations done in the following section, using a lack of progress in the (observed) log-likelihood yields good results. Thus, we consider the EM to have converged when there is a lack of progress in the log-likelihood, i.e., when
\begin{equation}
l^{(k+1)} - l^{(k)} < \epsilon,
\end{equation}
where $l^{(k)}$ is the log-likelihood value at iteration $k$.

\subsection{Model Selection}
After fitting the model for different values of $G$ and $q$, the Bayesian information criterion \citep[BIC;][]{schwarz78}  is used to select the ``best'' model amongst those fitted. It is given by
\begin{equation}
\label{eq:BIC}
\text{BIC} = 2 l(\bm{\hat\vartheta}) - \rho \, \text{log} \, n ,
\end{equation}
where $\bm{\hat\vartheta}$ is the maximum likelihood estimate of $\bm{\vartheta}$, 
$l(\bm{\hat\vartheta})$ is the maximized log-likelihood, $\rho$ is the number of free parameters, and $n$ is the number of observations.
Because $l(\bm{\hat\vartheta})$ surely increases as the number of parameters increases, the second term in (\ref{eq:BIC}) provides a penalty for the number of parameters.
Note that, assuming equal prior probabilities, the difference in BIC values between two models approximates the Bayes factor for a test concerning those models \citep{dasgupta98}. The fitted model with the largest BIC value is selected as the best model.


\section{Illustrations}
\subsection{Performance Assessment}
The Rand index \citep[RI;][]{rand71} can be used to assess classification performance. It is given by
\begin{equation}
\text{RI} = \frac{\text{number of pairwise agreements}}{\text{total number of pairs}},
\label{eq:AR}
\end{equation}
where the number of pairwise agreements is equal to the number of pairs of observations that belong to the same class and are grouped as such, plus the number of pairs that belong to separate classes and are grouped as such; the total number of pairs is $N \choose {2}$.
The adjusted Rand index \citep[ARI;][]{hubert85} corrects for chance agreement between pairs that tends to inflate the value in (\ref{eq:AR}). 
The ARI has an expected value of 0 under random classification and a value of 1 under perfect classification.
Clustering can be performed on data where group memberships are known by treating the data as unlabelled, and the resulting ARI value can then be used to assess the clustering performance of the model.

\subsection{Simulated Data Set 1}
\begin{figure}[!h]
\centering
\includegraphics[width=0.75\textwidth,height=0.45\textwidth]{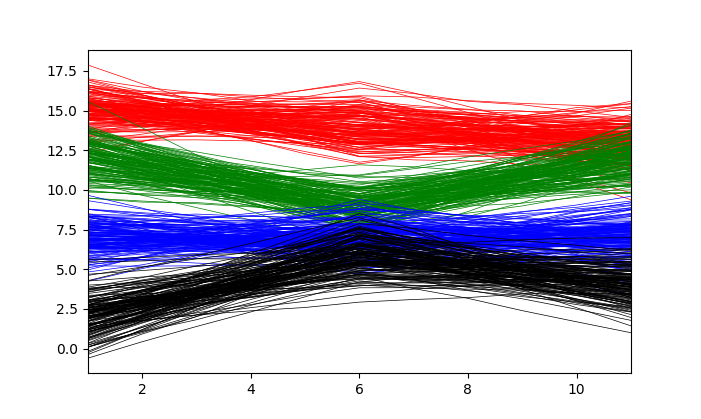} 
\caption{Simulated Dataset 1. The component memberships are indicated by colour.}\label{fig:simdata1pic}
\end{figure}

To illustrate the performance of the proposed model, the EM algorithm outlined in Section \ref{sec:modelfitting} is run on simulated data and the resulting clustering performance is assessed. Data are generated from $G=4$ distributions, where $n_1=n_2=n_3=n_4=150$. Each observation consists of $p=11$ observed time points which are generated from $3$-dimensional latent mean vectors, given by, respectively, $\xi_1=(2,6,4)'$, $\xi_2=(7,7,7)'$, $\xi_3=(12,9,12)'$, and $\xi_4=(15,14,13)'$. The matrix of factor loadings is given by
\[ \bLambda = \left( \begin{array}{ccccccccccc}
1.0 & 0.8 & 0.6 & 0.4 & 0.2 & 0.0 & 0.0 & 0.0 & 0.0 & 0.0 & 0.0 \\
0.0 & 0.2 & 0.4 & 0.6 & 0.8 & 1.0 & 0.8 & 0.6 & 0.4 & 0.2 & 0.0 \\
0.0 & 0.0 & 0.0 & 0.0 & 0.0 & 0.0 & 0.2 & 0.4 & 0.6 & 0.8 & 1.0 \end{array} \right)',\]
and the resulting data are shown in Figure \ref{fig:simdata1pic}.
The EM algorithm is then run for $G=1,\dots,6$ components and $q=2, 3, 4$ latent time points. For each $G$ and $q$, multiple starting values, using both random starting values and $k$-means clustering, are used to initialize the $z_{ig}$.

The model chosen by the BIC is indeed the ``correct" model, in that $G=4$ and $q=3$. This model classifies all the observations correctly, as seen in Table \ref{table:classification1}, and so the associated ARI is~1. 
The fitted model consists of 74 parameters in total. When comparing this to the standard (unconstrained) Gaussian mixture model that does not utilize any latent factors, we find that, by setting $G=4$ and $p=11$ in 
\begin{equation}
(G-1) + Gp + G \left[ \frac{p (p-1)}{2} \right] + Gp,
\label{eq:parameters2}
\end{equation} clustering would require 311 parameters.
\begin{table}[!h]
\caption{True component memberships cross-tabulated against predicted group memberships for the first simulated data set.}
\centering\begin{tabular*}{0.75\textwidth}{@{\extracolsep{\fill}}lcccr}
\hline
& \multicolumn{4}{c}{Predicted}\\
\cline{2-5}	
& Group 1 & Group 2 & Group 3 & Group 4\\ 
 \hline  
Component 1 & 150 & 0 & 0 & 0\\
Component 2 & 0 & 150 & 0 & 0\\
Component 3 & 0 & 0 & 150 & 0\\
Component 4 & 0 & 0 & 0 & 150\\
\hline
\end{tabular*}
\label{table:classification1}
\end{table}

\subsection{Simulated Data Set 2}
\begin{figure}[!ht]
\centering
\includegraphics[width=0.75\textwidth,height=0.45\textwidth]{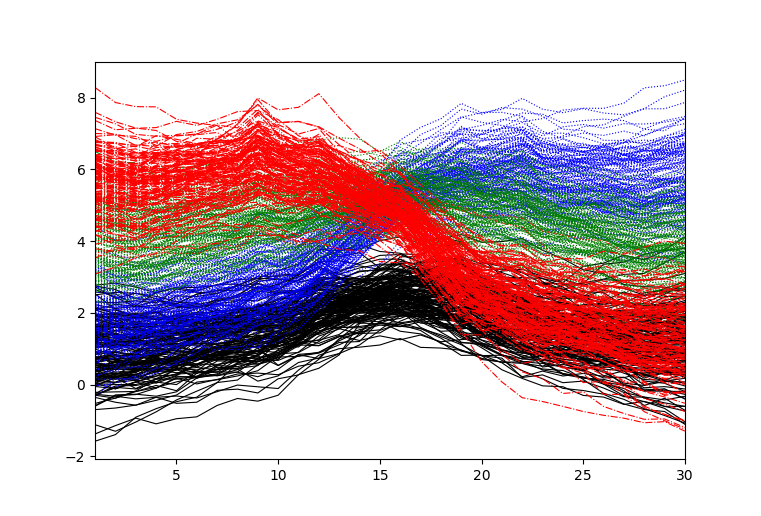} 
\caption{Simulated Dataset 2. The true component memberships are indicated by colours and the estimated group memberships are indicated by line style: alternating dashes, small dots, and solid lines represent Groups 1, 2, and 3, respectively.}\label{fig:simdata2pic}
\end{figure}

For the second simulation study, data are again generated from $G=4$ distributions, where $n_1=n_2=n_3=n_4=150$. This time, each observation consists of $p=30$ observed time points which are generated from $7$-dimensional latent mean vectors, given by, respectively, $\xi_1=(6,5,3,3,2,2,1.5)' $, $\xi_2=(4.5,3,3.5,2,3,3.5,4)'$, $\xi_3=(2,2,2.5,3,3.4,6)'$, and $\xi_4=(1,1,2,2,2,1,1)'$; the matrix of factor loadings is a $30 \times 7$ matrix. The data are shown in Figure \ref{fig:simdata2pic}. Note that this clustering problem is somewhat more challenging than the previous one. 

\begin{figure}[!ht]
\centering
\includegraphics[width=0.75\textwidth,height=0.45\textwidth]{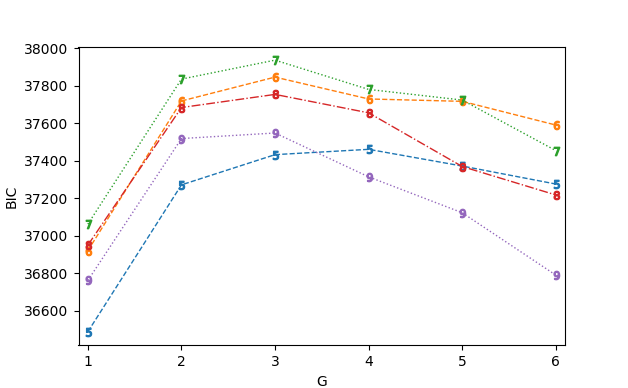}
\caption{BIC values versus the number of groups, where the numbers $5,\dots,9$ indicate the number of latent time points. }
\label{fig:simdata2BIC}
\end{figure}

The EM algorithm is run for $G=1,\dots,6$ components and $q=5,\dots,9$ latent time points, again using multiple runs of both random starting values and $k$-means clustering to initialize the $\hat{z}_{ig}$ values. The resulting BIC values are shown in Figure \ref{fig:simdata2BIC}: two, three, or perhaps four groups, and seven time latent time points, appear to give the best fit for the data. The best model has $G=3$ components and $q=7$ latent time points, and the resulting clustering performance of this model is summarized in Table \ref{table:classification2}. The associated ARI is 0.69, indicating good clustering performance despite the ``correct" number of components not being chosen. 
More specifically, we see from Table~\ref{table:classification2} that observations from components 1 and 4 are almost perfectly classified, while observations from components 2 and 3 are essentially classified into one group.
\begin{table}[!ht]
\caption{True group memberships (A--D) cross-tabulated against predicted group memberships (1--3) for the second simulated data set. 
}
\centering\begin{tabular*}{0.75\textwidth}{@{\extracolsep{\fill}}lccr}
\hline
& 1 & 2 & 3\\ 
 \hline  
A & 150 & 0 & 0\\
B & 0 & 149 & 1\\
C & 0 & 150 & 0\\
D & 0 & 4 & 146\\
\hline
\end{tabular*}
\label{table:classification2}
\end{table}

A total of 298 parameters are required for our fitted model. Without utilizing any latent factors, the (unconstrained) component covariance matrices alone would require $G[p(p-1)/2]+Gp = 465  G$ parameters. 
Thus, the model was able to achieve good clustering performance on this longitudinal data set using a significantly reduced number of parameters.


\subsection{Yeast Sporulation Data Set}
The final data set consists of 6,118 expressions of yeast genes measured at $p=7$ time points, analyzed previously by various researchers to find groups of genes exhibiting similar expression patterns \cite[]{mitchell94,chu98,wakefield03, mcnicholas10b}.
\begin{figure}[!ht]
\centering
\includegraphics[width=0.85\textwidth,height=0.5\textwidth]{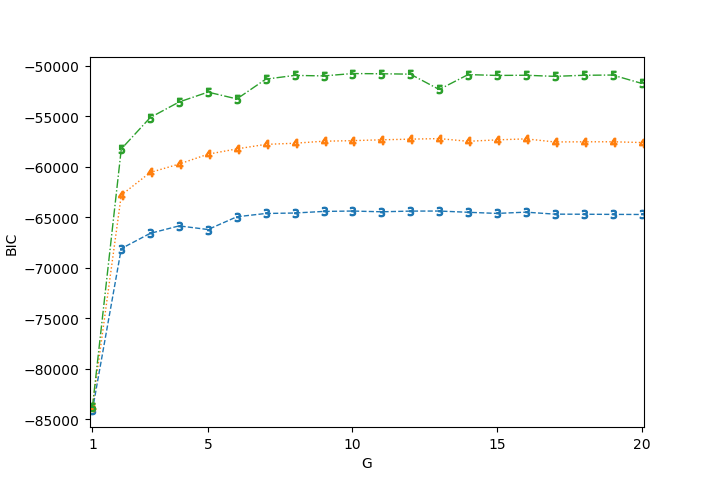}
\caption{BIC values versus the number of groups, where the numbers 3, 4, and 5 indicate the number of latent time points.}
\label{fig:yeastBIC}
\end{figure}
\begin{figure}[!ht]
\centering
\includegraphics[width=0.85\textwidth,height=0.5\textwidth]{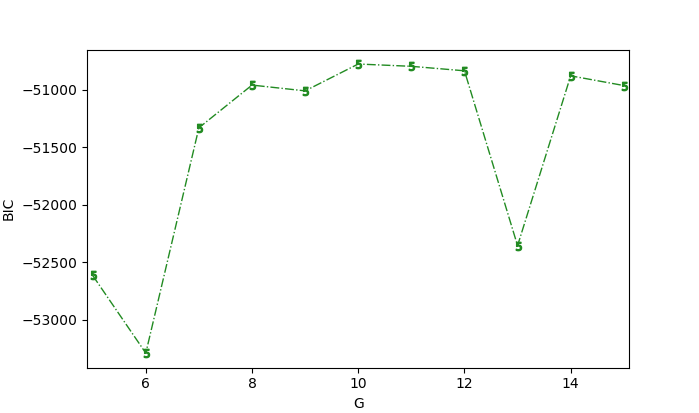}
\caption{BIC values versus the number of groups for $q=5$ latent time points.}
\label{fig:yeastBIC2}
\end{figure}
\begin{figure}[!ht]
\centering
\includegraphics[width=0.85\textwidth,height=0.5\textwidth]{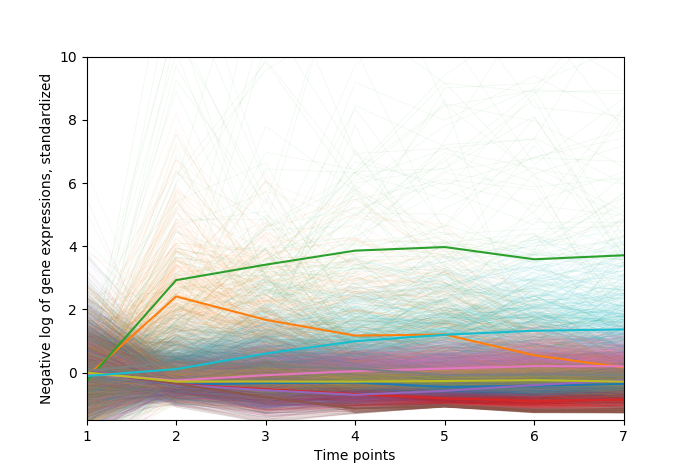}
\caption{Yeast gene expressions (after taking the negative logarithm, base 2, and standardizing) coloured by estimated group memberships, with estimated group mean trajectories (given by $\hat\bLambda \hat{\bxi}_g$ for $g=1,\dots,10$).}
\label{fig:yeastplot}
\end{figure}
To test the performance of the proposed model, we follow the analysis of \cite{mcnicholas10b} and consider all seven time points without the deletion of any genes, and take the negative logarithm, base 2, of each observation prior to analysis. The observations at each time point are also standardized to have mean 0 and variance 1.
The model is then fitted for $G=1,\dots,20$ components and $q=3,4,5$ latent time points. The $z_{ig}$ are again initialized using both random starting values and $k$-means clustering. The resulting BIC values for each model are shown in Figure \ref{fig:yeastBIC}. The best model, as seen more clearly in Figure \ref{fig:yeastBIC2}, has $G$ = 10 groups and $q$ = 5 latent time points. The resulting groups are pictured in Figure \ref{fig:yeastplot}, and are compared to the groups found by \cite{mcnicholas10b} in Table \ref{table:classification3}.
\begin{table}[!ht]
\centering
\caption{Predicted group memberships (A--M) found by \cite{mcnicholas10b} cross-tabulated against predicted group memberships found by the new model (1--10), for the yeast data set.}
\label{table:classification3}
\begin{tabular}{lllllllllll}
\hline
         & \multicolumn{10}{c}{Predicted}                          \\
         \cline{2-11}
         & 1   & 2   & 3  & 4   & 5   & 6  & 7   & 8   & 9   & 10  \\
\hline
A \hspace{5mm}  & 15  & 0   & 0  & 0   & 0   & 0  & 85  & 15  & 7   & 28  \\
B  & 7   & 23  & 0  & 0   & 13  & 0  & 439 & 131 & 175 & 160 \\
C  & 56  & 2   & 0  & 57  & 325 & 0  & 54  & 31  & 5   & 9   \\
D  & 640 & 0   & 0  & 77  & 120 & 0  & 460 & 196 & 765 & 11  \\
E  & 41  & 0   & 0  & 0   & 0   & 0  & 8   & 0   & 188 & 0   \\
F  & 3   & 0   & 1  & 21  & 3   & 89 & 1   & 10  & 1   & 0   \\
G  & 0   & 97  & 4  & 0   & 0   & 0  & 0   & 1   & 0   & 2   \\
H  & 0   & 5   & 60 & 0   & 2   & 0  & 22  & 10  & 0   & 165 \\
I  & 1   & 23  & 17 & 0   & 1   & 0  & 91  & 10  & 25  & 107 \\
J & 20  & 15  & 11 & 41  & 30  & 47 & 14  & 13  & 7   & 11  \\
K & 94  & 10  & 0  & 326 & 12  & 26 & 11  & 138 & 16  & 8   \\
L & 50  & 0   & 0  & 63  & 3   & 8  & 6   & 3   & 13  & 0   \\
M & 0   & 127 & 18 & 0   & 0   & 0  & 3   & 49  & 0   & 10 \\
\hline
\end{tabular}
\end{table}

From Table \ref{table:classification3}, it seems clear that the model introduced herein found different groups of gene expressions.
%
The fitted model requires 226 parameters, while the model that does not utilize any latent factors would require 359 parameters.

\section{Summary}
A mixture model for clustering high-dimensional longitudinal data was introduced, which has the ability to account for the longitudinal nature of the data being clustered while also providing significant data reduction. This is achieved using an extension of the mixture of common factor analyzers model, whereby the variability in measurements taken at many time points can be explained using a smaller number of (latent) time points. A modified Cholesky decomposition is used on the latent covariance matrices to account for the relationship between measurements in the latent space. Constraints can be placed on these latent covariance matrices to further reduce the number of parameters; however, the mixture of common factor analyzers approach used in this model provides significant data reduction even without doing so.

The model was fit to two simulated data sets and one real data set, overall demonstrating good clustering performance using a significantly reduced number of parameters. Use of the model on the first simulated data set resulted in a perfect clustering result, and use of the model on the second simulated data set, which consisted of 30 time points, yielded good clustering performance using only seven latent time points. Use of the model on the yeast sporulation data set allowed clustering of gene expressions to be achieved using a reduced number of parameters.

Further work could include applications to (real) longitudinal data sets consisting of a large number of time points. As well, note that the model outlined here is only useful for clustering univariate longitudinal data\textemdash that is, data that contain only one measurement taken at each time point. \cite{anderlucci14} use a mixture of matrix variate normal distributions to cluster multivariate longitudinal data; thus, future work could involve extending the model introduced here to account for longitudinal data that contain multiple measurements taken at each time point. Further, this model may be extended to include non-Gaussian distributions; for example, \cite{gallaugher2017b} detail clustering using mixtures of skewed matrix variate distributions.


\appendix
\section{Derivations of parameter estimates}\label{sec:appendix}
Differentiating $Q$, given by (\ref{eq:Q}), with respect to $\bxi_g$, $\bT_g$, $\bD_g^{-1}$, $\bLambda$, and $\bPsi^{-1}$, respectively, gives the score functions
\begin{align*}
S_{1}(\bm{\vartheta}_{s}) = \frac{\partial Q}{\partial \bxi_g} 
&= - \frac{1}{2}  \sum_{i=1}^{n} \hat{z}_{ig} \frac{\partial}{\partial \bxi_g} \text{tr} \left\{ \bT_{g}^{'} \bD_g^{-1} \bT_{g}
\mathbb{E} \lbrack (\mathbf{U}_{ig}-\bxi_g)(\mathbf{U}_{ig}-\bxi_g)^{'} \mid \mathbf{x}_{i}, z_{ig} = 1 \rbrack \right\}
\\ &= \frac{1}{2}  \sum_{i=1}^{n} \hat{z}_{ig} \Big[ \bT_{g}^{'} \bD_g^{-1} \bT_{g} \mathbb{E} \lbrack \mathbf{U}_{ig} \mid \mathbf{x}_{i}, z_{ig} = 1 \rbrack + \bT_{g}^{'} \bD_g^{-1} \bT_{g} \mathbb{E} \lbrack \mathbf{U}_{ig} \mid \mathbf{x}_{i}, z_{ig} = 1 \rbrack 
\\& \hspace{3em} - (\bT_{g}^{'} \bD_g^{-1} \bT_{g} + \bT_{g}^{'} \bD_g^{-1} \bT_{g} ) \bxi_g \Big]
\\ &= \bT_{g}^{'} \bD_g^{-1} \bT_{g} \sum_{i=1}^{n} \hat{z}_{ig} \left( \mathbb{E} \lbrack \mathbf{U}_{ig} \mid \mathbf{x}_{i}, z_{ig} = 1 \rbrack - \bxi_g \right),
\\S_{2}(\bm{\vartheta}_{s}) = \frac{\partial Q}{\partial \bT_{g}}
&= - \frac{1}{2}  \sum_{i=1}^{n} \hat{z}_{ig} \frac{\partial}{\partial \bT_{g}} \text{tr} \left\{ \bT_{g}^{'} \bD_g^{-1} \bT_{g}
\mathbb{E} \lbrack (\bU_{ig}-\bxi_g)(\mathbf{U}_{ig}-\bxi_g)^{'} \mid \mathbf{x}_{i}, z_{ig} = 1 \rbrack \right\}
\\ &= - n_{g} \bD_g^{-1} \bT_{g} \frac{1}{n_{g}} \sum_{i=1}^{n} \hat{z}_{ig} \mathbb{E} \lbrack (\bU_{ig}-\bxi_g)(\bU_{ig}-\bxi_g)^{'} \mid \mathbf{x}_{i}, z_{ig} = 1 \rbrack
\\ &= - n_{g} \bD_g^{-1} \bT_{g} \mathbf{S}_{g}, 
\end{align*} where
\begin{align*}
\mathbf{S}_{g} &= \frac{1}{n_{g}} \sum_{i=1}^{n} \hat{z}_{ig}
\mathbb{E} \lbrack (\bU_{ig}-\bxi_g)(\bU_{ig}-\bxi_g)^{'} \mid \mathbf{x}_{i}, z_{ig} = 1 \rbrack
\\ & = \frac{1}{n_{g}} \sum_{i=1}^{n} \hat{z}_{ig} \left( \mathbb{V}\text{ar} \lbrack \bU_{ig} - \bxi_g \mid \mathbf{x}_{i}, z_{ig} = 1 \rbrack + \mathbb{E} \lbrack \bU_{ig} - \bxi_g \mid \mathbf{x}_{i}, z_{ig} = 1 \rbrack \mathbb{E} \lbrack (\bU_{ig} - \bxi_g)^{'} \mid \mathbf{x}_{i}, z_{ig} = 1 \rbrack \right)
\\& = \frac{1}{n_{g}} \sum_{i=1}^{n} \hat{z}_{ig} \left[ (\mathbf{I}_{q}-\hat{\bBeta}_{g} \hat{\bLambda}) (\hat{\bT}_{g}^{'} \hat{\bD}_g^{-1} \hat{\bT}_{g})^{-1} + \hat{\bBeta}_{g} ( \mathbf{x}_{i} - \hat{\bLambda} \hat{\bxi}_{g}) ( \mathbf{x}_{i} - \hat{\bLambda} \hat{\bxi}_{g})^{'} \hat{\bBeta}_{g}^{'} \right],
\end{align*}
\begin{align*}
S_{3}(\bm{\vartheta}_{s}) = \frac{\partial Q}{\partial \bD_g^{-1}} 
&= \frac{1}{2} \sum_{i=1}^{n} \hat{z}_{ig} \Bigg[ \frac{1}{\left| \bD_g^{-1}\right|} \frac{\partial}{\partial \bD_g^{-1}} \left| \bD_g^{-1} \right| \\ & \hspace{3em} - \frac{\partial}{\partial \bD_g^{-1}} \text{tr} \left\{ \bT_{g}^{'} \bD_g^{-1} \bT_{g} \mathbb{E} \lbrack (\bU_{ig}-\bxi_g)(\bU_{ig}-\bxi_g)^{'} \mid \mathbf{x}_{i}, z_{ig} = 1 \rbrack \right\} \Bigg]
\\ &= \frac{n_{g}}{2} \bD_g - \frac{1}{2} \sum_{i=1}^{n} \hat{z}_{ig} \bT_{g} \Big( \mathbb{V}\text{ar} \lbrack \bU_{ig} - \bxi_g \mid \mathbf{x}_{i}, z_{ig} = 1 \rbrack \\& \hspace{3em}
+ \mathbb{E} \lbrack \bU_{ig} - \bxi_g \mid \mathbf{x}_{i}, z_{ig} = 1 \rbrack \mathbb{E} \lbrack (\bU_{ig} - \bxi_g)^{'} \mid \mathbf{x}_{i}, z_{ig} = 1 \rbrack \Big) \bT_{g}^{'},
\end{align*}
\begin{align*}
S_{4}(\bm{\vartheta}_{s}) = \frac{\partial Q}{\partial \bLambda} 
&= - \frac{1}{2}  \sum_{i=1}^{n} \sum_{g=1}^{G} \hat{z}_{ig} \frac{\partial}{\partial \bLambda} \text{tr} \left\{ \bPsi^{-1} \mathbb{E} \lbrack (\mathbf{x}_{i} - \bLambda \bU_{ig}) (\mathbf{x}_{i} - \bLambda \bU_{ig})^{'} \mid \mathbf{x}_{i}, z_{ig} = 1 \rbrack \right\}
\\& = - \frac{1}{2}  \sum_{i=1}^{n} \sum_{g=1}^{G} \hat{z}_{ig} \frac{\partial}{\partial \bLambda} \text{tr} \Big\{ \bPsi^{-1} \Big( \mathbf{x}_{i} \mathbf{x}_{i}^{'} - \bLambda \mathbb{E} \left[ \bU_{ig} \mid \mathbf{x}_{i}, z_{ig} = 1 \right] \mathbf{x}_{i}^{'} - \mathbf{x}_{i} \mathbb{E} \left[ \bU_{ig}^{'} \mid \mathbf{x}_{i}, z_{ig} = 1 \right] \bLambda^{'} \\& \hspace{3em}
+ \bLambda \mathbb{E} \left[ \bU_{ig} \bU_{ig}^{'} \mid \mathbf{x}_{i}, z_{ig} = 1 \right] \bLambda^{'} \Big) \Big\}
\\& = \sum_{i=1}^{n} \sum_{g=1}^{G} \hat{z}_{ig} \bPsi^{-1} \left( \mathbf{x}_{i} \mathbb{E} \lbrack \bU_{ig}^{'} \mid \mathbf{x}_{i}, z_{ig} = 1 \rbrack - \bLambda \mathbb{E} \lbrack \bU_{ig} \bU_{ig}^{'} \mid \mathbf{x}_{i}, z_{ig} = 1 \rbrack \right),
\end{align*}
\begin{align*}
S_{5}(\bm{\vartheta}_{s}) = \frac{\partial Q}{\partial \bPsi^{-1}} 
&= \frac{1}{2} \sum_{i=1}^{n} \sum_{g=1}^{G} \hat{z}_{ig}
\bigg[ \frac{1}{\left| \bPsi^{-1} \right|} \frac{\partial}{\partial \bPsi^{-1}} \left| \bPsi^{-1} \right| \\& \hspace{3em} - \frac{\partial}{\partial \bPsi^{-1}} \text{tr} \left \{ \bPsi^{-1} \mathbb{E} \lbrack (\mathbf{x}_{i} - \bLambda \bU_{ig}) (\mathbf{x}_{i} - \bLambda \bU_{ig})^{'} \mid \mathbf{x}_{i}, z_{ig} = 1 \rbrack \right \} \bigg]
\\& = \frac{1}{2} \sum_{i=1}^{n} \sum_{g=1}^{G} \hat{z}_{ig}
\left( \bPsi - \mathbb{E} \lbrack (\mathbf{x}_{i} - \bLambda \bU_{ig}) (\mathbf{x}_{i} - \bLambda \bU_{ig})^{'} \mid \mathbf{x}_{i}, z_{ig} = 1 \rbrack \right)
\\& = \frac{n}{2} \bPsi - \frac{1}{2} \sum_{i=1}^{n} \sum_{g=1}^{G} \hat{z}_{ig} \Big( \mathbf{x}_{i} \mathbf{x}_{i}^{'} - \bLambda \mathbb{E} \lbrack \bU_{ig} \mid \mathbf{x}_{i}, z_{ig} = 1 \rbrack \mathbf{x}_{i}^{'} - \mathbf{x}_{i} \mathbb{E} \lbrack \bU_{ig}^{'} \mid \mathbf{x}_{i}, z_{ig} = 1 \rbrack \bLambda^{'} \\& \hspace{3em}
+ \bLambda \mathbb{E} \lbrack \bU_{ig} \bU_{ig}^{'} \mid \mathbf{x}_{i}, z_{ig} = 1 \rbrack \bLambda^{'} \Big). \\
\end{align*}


Solving
$S_1 (\hat{\bxi}_{g}^{\text{new}}, \hat{\bT}_{g}^{\text{new}}, \hat{\bD}_{g}^{\text{new}}, \hat{\bLambda}^{\text{new}}, \hat{\bPsi}^{\text{new}}) = \bf{0}$ gives
\begin{align*}
&\hat{\bT}_{g}^{'} \hat{\bD}_{g}^{-1} \hat{\bT}_{g} \sum_{i=1}^{n} \hat{z}_{ig} \left( \mathbb{E} \lbrack \mathbf{U}_{ig} \mid \mathbf{x}_{i}, z_{ig} = 1 \rbrack - \hat{\bxi}_{g}^{\text{new}} \right) = \bf{0}
\\ & \implies n_{g} \hat{\bxi}_{g}^{\text{new}} = \sum_{i=1}^{n} \hat{z}_{ig} \mathbb{E} \lbrack \mathbf{U}_{ig} \mid \mathbf{x}_{i}, z_{ig} = 1 \rbrack
\\ & \implies \hat{\bxi}_{g}^{\text{new}} = \hat{\bxi}_g + \frac{1}{n_g} \sum_{i=1}^{n} \hat{z}_{ig} \hat{\Beta}_{g} (\mathbf{x}_{i} - \hat{\bLambda} \hat{\bxi}_{g} ).
\end{align*}
Then, solving
$\text{diag}\left\{ S_3 (\hat{\bxi}_{g}^{\text{new}}, \hat{\bT}_{g}^{\text{new}}, \hat{\bD}_g^{\text{new}}, \hat{\bLambda}^{\text{new}}, \hat{\bPsi}^{\text{new}}) \right\} = \bf{0}$,
\begin{align*}
&\frac{n_{g}}{2} \hat{\bD}_g^{\text{new}} - \frac{1}{2} \text{diag} \Bigg\{ \sum_{i=1}^{n} \hat{z}_{ig} \hat{\bT}_{g}  \Big( \mathbb{V}\text{ar} \lbrack \bU_{ig} - \bxi_g \mid \mathbf{x}_{i}, z_{ig} = 1 \rbrack \\&\hspace{11em} + \mathbb{E} \lbrack \bU_{ig} - \bxi_g \mid \mathbf{x}_{i}, z_{ig} = 1 \rbrack \mathbb{E} \lbrack (\bU_{ig} - \bxi_g)^{'} \mid \mathbf{x}_{i}, z_{ig} = 1 \rbrack \Big) \hat{\bT}_{g}^{'} \Bigg\} = \bf{0}
\\& \implies \hat{\bD}_g^{\text{new}} = \text{diag} \left\{ \hat{\bT}_{g} (\mathbf{I}_{q}-\hat{\bBeta}_{g} \hat{\bLambda}) 
\hat{\bT}_{g}^{-1} \hat{\bD}_g + \frac{1}{n_{g}} \sum_{i=1}^{n} \hat{z}_{ig} \hat{\bT}_{g} \hat{\bBeta}_{g} ( \mathbf{x}_{i} - \hat{\bLambda} \hat{\bxi}_{g}) ( \mathbf{x}_{i} - \hat{\bLambda} \hat{\bxi}_{g})^{'} \hat{\bBeta}_{g}^{'} \hat{\bT}_{g}^{'} \right\},
\end{align*}
and
 $S_4 (\hat{\bxi}_{g}^{\text{new}}, \hat{\bT}_{g}^{\text{new}}, \hat{\bD}_g^{\text{new}}, \hat{\bLambda}^{\text{new}}, \hat{\bPsi}^{\text{new}}) = \bf{0}$
\begin{align*}
 & \implies \sum_{i=1}^{n} \sum_{g=1}^{G} \hat{z}_{ig} \bPsi^{-1} \left( \mathbf{x}_{i} \mathbb{E} \lbrack \bU_{ig}^{'} \mid \mathbf{x}_{i}, z_{ig} = 1 \rbrack - \bLambda \mathbb{E} \lbrack \bU_{ig} \bU_{ig}^{'} \mid \mathbf{x}_{i}, z_{ig} = 1 \rbrack \right) = \bf{0}
\\ & \implies \hat{\bLambda}^{\text{new}} = \left\{
\sum_{i=1}^{n} \sum_{g=1}^{G} \hat{z}_{ig} \mathbf{x}_{i} \left[
\hat{\bxi}_g + \hat{\Beta}_{g} (\mathbf{x}_{i} - \hat{\bLambda} \hat{\bxi}_{g} )
\right]^{'} \right \} \Bigg\{ \sum_{i=1}^{n} \sum_{g=1}^{G} \hat{z}_{ig} (\mathbf{I}_{q} - \hat{\Beta}_{g} \hat{\bLambda}) (\hat{\bT}_{g}^{'} \hat{\bD}_g^{-1} \hat{\bT}_{g})^{-1} \\& \hspace{10em}
+ \sum_{i=1}^{n} \sum_{g=1}^{G} \hat{z}_{ig} \lbrack \hat{\bxi}_{g} + \hat{\bBeta}_{g} ( \mathbf{x}_{i} - \hat{\bLambda} \hat{\bxi}_{g}) \rbrack \lbrack \hat{\bxi}_{g} + \hat{\bBeta}_{g} ( \mathbf{x}_{i} - \hat{\bLambda} \hat{\bxi}_{g}) \rbrack ^{'} \Bigg\}^{-1}.
\end{align*}
Finally, 
 $\text{diag} \Big\{ S_5 (\hat{\bxi}_{g}^{\text{new}}, \hat{\bT}_{g}^{\text{new}}, \hat{\bD}_g^{\text{new}}, \hat{\bLambda}^{\text{new}}, \hat{\bPsi}^{\text{new}}) \Big\} = \bf{0}$
\begin{align*}
& \implies \frac{n}{2} \bPsi^{\text{new}} - \frac{1}{2} \text{diag} \Bigg\{ \sum_{i=1}^{n} \sum_{g=1}^{G} \hat{z}_{ig} \Big( \mathbf{x}_{i} \mathbf{x}_{i}^{'} - \bLambda \mathbb{E} \lbrack \bU_{ig} \mid \mathbf{x}_{i}, z_{ig} = 1 \rbrack \mathbf{x}_{i}^{'} - \mathbf{x}_{i} \mathbb{E} \lbrack \bU_{ig}^{'} \mid \mathbf{x}_{i}, z_{ig} = 1 \rbrack \bLambda^{'} \\& \hspace{10em}
+ \bLambda \mathbb{E} \lbrack \bU_{ig} \bU_{ig}^{'} \mid \mathbf{x}_{i}, z_{ig} = 1 \rbrack \bLambda^{'} \Big) \Bigg\} = \bf{0}
\\& \implies \hat{\bPsi}^{\text{new}} = \frac{1}{n} \text{diag}  \Bigg\{ \sum_{i=1}^{n} \sum_{g=1}^{G} \hat{z}_{ig} \Big( \mathbf{x}_{i} \mathbf{x}_{i}^{'} - \hat{\bLambda} \mathbb{E} \lbrack \bU_{ig} \mid \mathbf{x}_{i}, z_{ig} = 1 \rbrack \mathbf{x}_{i}^{'} - \mathbf{x}_{i} \mathbb{E} \lbrack \bU_{ig}^{'} \mid \mathbf{x}_{i}, z_{ig} = 1 \rbrack \hat{\bLambda}^{'} \\& \hspace{10em}
+ \hat{\bLambda} \mathbb{E} \lbrack \bU_{ig} \bU_{ig}^{'} \mid \mathbf{x}_{i}, z_{ig} = 1 \rbrack \hat{\bLambda}^{'} \Big) \Bigg\}
\\& \implies \hat{\bPsi}^{\text{new}} = \frac{1}{n} \text{diag} \Bigg\{ \sum_{i=1}^{n} \sum_{g=1}^{G} \hat{z}_{ig}
\bigg( \mathbf{x}_{i} \mathbf{x}_{i}^{'} - \hat{\bLambda} \left[ \hat{\bxi}_g + \hat{\Beta}_{g} (\mathbf{x}_{i} - \hat{\bLambda} \hat{\bxi}_{g} ) \right] \mathbf{x}_{i}^{'} \\& \hspace{10em}  - \mathbf{x}_{i} \left[ \hat{\bxi}_g + \hat{\Beta}_{g} (\mathbf{x}_{i} - \hat{\bLambda} \hat{\bxi}_{g} ) \right] ^{'} \hat{\bLambda}^{'} + \hat{\bLambda} \bigg[ (\mathbf{I}_{q} - \hat{\Beta}_{g} \hat{\bLambda})(\hat{\bT}_{g}^{'} \hat{\bD}_g^{-1} \hat{\bT}_{g} )^{-1} \\& \hspace{10em} + \Big[ \hat{\bxi}_g + \hat{\Beta}_{g} (\mathbf{x}_{i} - \hat{\bLambda} \hat{\bxi}_{g} ) \Big] \Big[ \hat{\bxi}_g + \hat{\Beta}_{g} (\mathbf{x}_{i} - \hat{\bLambda} \hat{\bxi}_{g} ) \Big] ^{'} \bigg] \hat{\bLambda}^{'} \bigg) \Bigg\}.
\end{align*}

{\small

}
\end{document}